\documentclass[%
 reprint,
 amsmath,amssymb,
 aps,
 pdflatex,
]{revtex4-2}

\usepackage{graphicx}
\usepackage{dcolumn}
\usepackage{bm}
\usepackage{physics}
\usepackage{xcolor}
\usepackage{geometry}
\usepackage{caption}
\usepackage{subfiles}
\usepackage{lineno}
\usepackage{ulem}
\usepackage[utf8]{inputenc}
\usepackage{xr}
\usepackage{graphicx}
\newcommand{\der}[2]{\frac{\partial}{\partial #2}#1}

\definecolor{orangered}{rgb}{1,0.4,0}
\definecolor{dred}{rgb}{0.83,0.15,0.15}
\definecolor{dgreen}{rgb}{0.033,0.35,0.033}
\definecolor{dyellow}{rgb}{0.45,0.45,0.0}
\definecolor{trq}{rgb}{0.05,0.45,0.45}
\definecolor{lblue}{rgb}{0.15,0.35,0.75}

\newcommand{\voc}{\rm{V_{oc}}}
\DeclareUnicodeCharacter{2215}{\slash}

\begin{document}

\title[Femto-to-Giga]{Hydrogen-induced degradation dynamics in silicon heterojunction solar cells via machine learning}

\author{Andrew \surname{Diggs}}
\email{amdiggs@ucdavis.edu}

\author{Zitong \surname{Zhao}}
\email{ztzhao@ucdavis.edu}

\author{Davis \surname{Unruh}}
\altaffiliation[Also at ]{Center for Nanoscale Materials, Argonne National Laboratory, Lemont, IL 60439, USA}

\author{Gergely T. \surname{Zim\'anyi}}

\affiliation{%
 Physics Department, University of California\\
 Davis, CA 95616, USA}%
 
\author{Reza Vatan \surname{Meidanshahi}}
\author{Salman \surname{Manzoor}}
\author{Mariana \surname{Bertoni}}
\author{Stephen M. \surname{Goodnick}}
\affiliation{
School of Electrical, Computer and Energy Engineering, Arizona State University\\ 
Tempe, AZ 85287, USA}




\begin{abstract}
Among silicon-based solar cells, heterojunction cells hold the world efficiency record. However, their market acceptance is hindered by an initial 0.5\% per year degradation of their open circuit voltage which doubles the overall cell degradation rate. Here, we study the performance degradation of crystalline-Si/amorphous-Si:H heterojunction stacks. First, we experimentally measure the interface defect density over a year, the primary driver of the degradation. Second, we develop SolDeg, a multiscale, hierarchical simulator to analyze this degradation by combining Machine Learning, Molecular Dynamics, Density Functional Theory, and Nudged Elastic Band methods with analytical modeling. We discover that the chemical potential for mobile hydrogen develops a gradient, forcing the hydrogen to drift from the interface, leaving behind recombination-active defects. We find quantitative correspondence between the calculated and experimentally determined defect generation dynamics. Finally, we propose a reversed Si-density gradient architecture for the amorphous-Si:H layer that promises to reduce the initial open circuit voltage degradation from 0.5\% per year to 0.1\% per year.  

\end{abstract}


\maketitle

 \section*{Introduction}
With record-breaking efficiencies approaching 27\%\cite{RecordEff}, silicon heterojunction (Si HJ) solar cells are rapidly becoming one of the most promising next generation technologies. A key driver for their excellent performance is the electronic passivation of the crystalline silicon (c-Si) by a thin layer of hydrogenated amorphous silicon (a-Si:H). In spite of their great promise, Si HJ solar cells' market acceptance is lagging, primarily because their efficiency degradation was reported to be about {0.7\% per yr\cite{NRELdegradation}, much higher than the usual rate of 0.2\% per yr. \cite{GAO2022}} The excess 0.5\% per yr rate was attributed to the degradation of the open circuit voltage $\voc$. \cite{NRELdegradation,GAO2022}  (It is noted that these are initial degradation rates, measured over limited time ranges. Nonlinear aspects over longer time ranges will be discussed below.) Therefore, understanding and subsequently minimizing the performance loss of Si HJ solar cells is key to accelerating their market acceptance. Towards this goal, this paper focuses on analyzing and potentially suppressing Si HJ solar cell degradation. For completeness, we add that their market is also limited by the need for disruptive changes in the manufacturing chain, increased cost of the n-type wafers, and the low-temperature Ag metallizations, among others.

Although a-Si:H has been analyzed for more than 40 years, several questions of its physics still remain open. Unresolved issues include understanding the classes of structural and electronic defects,\cite{staeblerwronski,JOHNSON1991Reviw,Shimizu2004Review,Wronski2014Review,Melskens2014,Guha1992,Stutzmann1985,Redfield1989} their formation and statistical distributions, and the long-term structural defect dynamics of a-Si:H. These issues all play crucial roles in the degradation of the crystalline/amorphous (c-Si/a-Si:H) interface of Si HJs. \cite{Bertoni2019,Holovsky2020,soldeg}

The primary form of degradation seen in Si HJ cells is an increase in surface recombination at the c-Si/a-Si:H interface. Here, dangling bonds (DBs) are known to be the primary recombination centers. \cite{DBrecombination} DBs are inherently present, created by disorder and thermodynamics. The recombination is suppressed by passivating these DBs with the introduction of hydrogen during fabrication. The degradation over the lifetime of the cell unfolds by the increase of the DB density. The DB density can increase by different mechanisms. In a previous paper we computed the increase of the DB density by the structural evolution of the Si matrix alone. \cite{soldeg} We found that the structural evolution of the Si matrix did increase the DB density, but quantitatively was not sufficient to account for the data. Therefore, in this paper we will focus on another mechanism that can drive the increase of the DB density: the loss of passivating hydrogen at the interface.

Over the past 30 years, many studies investigated hydrogen dynamics in bulk a-Si:H, focusing on its drift and diffusive motion and the formation of light induced defects (LID). \cite{BEYER1991,Beyer-1996,santos_johnson1993,Santos&Johnson-1992,streetkakalios,streetnostats,dispersive,biswas,VDW1994,van_de_walle_street_1995,ww,wronski1976,WRONSKI,Herring2001} Also, there have been numerous papers on the dynamics of hydrogen in a-Si:H and its role in passivating the surface of c-Si. \cite{BEYER1991,Beyer-1996,santos_johnson1993,Santos&Johnson-1992,streetkakalios,streetnostats,dispersive,biswas,VDW1994,van_de_walle_street_1995,ww,DBrecombination,DewolfBurrows2008,DeWolfKondo2007,Olibet2007,Bertoni2019} 

More recently, studies zoomed in on the dynamics at the c-Si/a-Si:H interface; specifically, on the Si-H bonding structure and restructuring during annealing and light soaking,\cite{Cattin,Veirman-2022,Olibet2006_LID} hydrogen motion at the interface at elevated temperatures,\cite{DeWolfSTREXP} the role of hydrogen in preventing epitaxial growth,\cite{Liu-2016,Sai-2018} and interface defect types and formation. \cite{DBrecombination,DewolfBurrows2008,DeWolfKondo2007,Olibet2007} These works on the kinetics of hydrogen in a-Si:H have provided crucial understanding of key aspects of the electronic passivation of c-Si. However, several issues remain open. Theoretically, the structural variability of amorphous Si necessitates the study of large samples which are hard to access by numerical methods. Experimentally, it has proven challenging to track hydrogen motion in silicon structures. The above challenges demonstrate the need for further theoretical and numerical research to provide additional insight. The most effective numerical methods are first principles electronic-structure studies using density functional theory (DFT). \cite{conceptual_DFT,herring,VDW1994,dftSWeffect,WRONSKI} However, the computational expense of DFT scales unfavorably with O(N$^3$). Thus, in order to reach larger sample sizes, many computational studies were performed with Molecular Dynamics (MD) methods using parameterized empirical interatomic potentials, whose cost scales linearly with O(N). \cite{mikefinnis2003} However, this gain in computational speed was achieved at the cost of precision. 

In the last decade the adoption of machine learning (ML) methods made it possible to create new interatomic potentials that improved MD methods to attain DFT level accuracy. \cite{GAP1, GAP2, PRX, Thompson2015, Chen2017, Behler2007, Behler2011, Behler2016, Kocer2019, ShapeevMTP2016, KresseVASPGAP, Tkatchenko2017, ANI, ACE} Motivated by these successes, a subset of the present authors very recently developed the first ML-trained Gaussian Approximation Potential (GAP) for hydrogenated silicon, the Si-H GAP. \cite{SiHGAP} It was demonstrated that the Si-H GAP was able to simulate samples of sizes unattainable for DFT with DFT level accuracy, as described in detail below.

Motivated by the above challenges, needs, and considerations, in this work we combine experimental and computational efforts to perform a comprehensive analysis of the degradation of c-Si/a-Si:H heterojunction systems. Experimentally, we have tracked the degradation in Si HJ systems over a year. Theoretically, we upgraded our SolDeg degradation simulator to track hydrogen-induced degradation in the same Si HJ systems from femtosecond to gigasecond time scales. Remarkably, we have achieved a compelling quantitative correspondence of our simulation results with the experimental data. Finally, building on the understanding developed with this SolDeg analysis, we developed an actionable proposition on how to modify the structure of Si HJ cells to dramatically suppress the  initial $\voc$ degradation  rate by as much as 80\%. Such a substantial reduction may considerably help the market acceptance of Si HJ cells.

\section*{Results and Discussion}
\subsection*{Experimental Degradation Study}

The goal of the experimental part of the project was to investigate the degradation dynamics of stacks of hydrogenated amorphous silicon, a-Si:H, deposited on top of crystalline silicon, c-Si. We created a set of n-type c-Si/a-Si:H stacks and measured the carrier lifetime over the period of a year under dark and ambient temperature conditions. The sample preparation and experimental techniques are described in the Methods section.

The effective minority carrier lifetime $( \tau_{\rm{eff}} )$ of a symmetrically passivated sample with low surface recombination velocity {(S)} is a combination of the individual lifetimes due to bulk, Auger, Shockley-Read-Hall (SRH), radiative, and surface recombination processes:

\begin{equation}
       \frac{1}{\tau_{\rm{eff}}} = \frac{1}{\tau_{aug}} + \frac{1}{\tau_{rad}} + \frac{1}{\tau_{SHR}} + \frac{ 2S}{W}
       \label{eqn:TauEff}
\end{equation}

where the surface recombination rate is the only one that depends of the wafer thickness $W$. We determined S from the slope of the $1/\tau_{\rm{eff}}$ versus $1/W$ plot, constructed from the lifetime data taken on four wafers with varying thicknesses $W$. The S was determined for a range of injection levels $\Delta n$ and a range of temperatures $T$ to construct  S$(\Delta n,T)$. Finally, we determined the defect density $N$ and the charge density $Q$ at the interface by fitting S$(\Delta n,T)$ to the amphoteric model proposed by Olibet \textit{et al.}. \cite{Olibet2007} We collected the S$(\Delta n, T)$ data every two weeks over the course of a year, and used these data to construct the time dependence of $N(t)$ and ${Q(t)}$.

\begin{figure}[h]
    \centering
    \includegraphics[width = 1.0\linewidth]{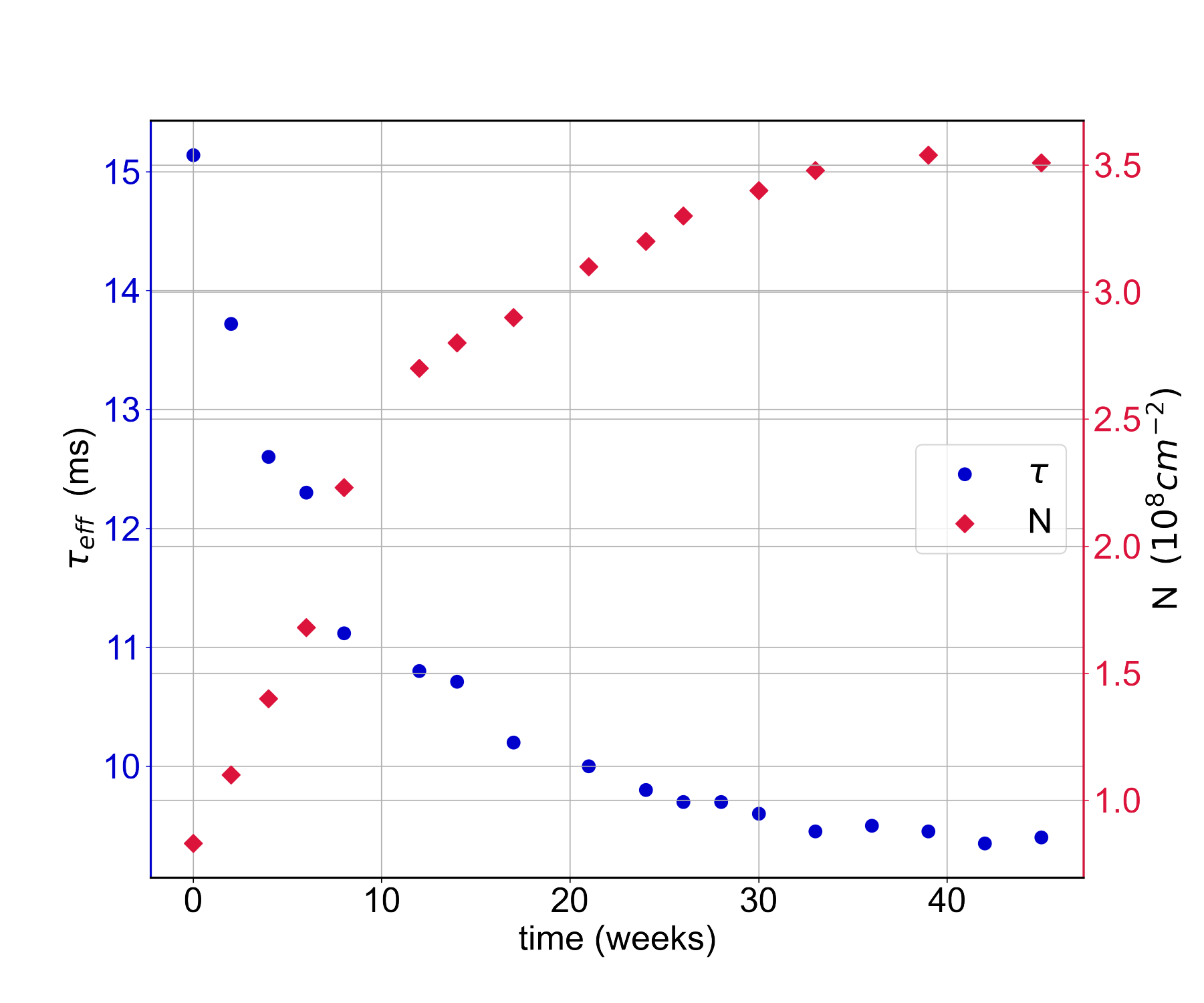}
     \caption{\textbf{Degradation of the surface passivation of silicon heterojunction solar cells.}  The quality of the surface passivation of four silicon heterojunction cells with varying thickness was tracked over the course of a year by monitoring the effective minority carrier lifetime $\tau_{\rm{eff}}$ (\textcolor{blue}{blue circles}). From analyzing these measurements at different temperatures and different injection levels, we determined the time-dependent defect density at the c-Si/a-Si:H interface $ \rm{N}(t)$ (\textcolor{dred}{red diamonds}).}
    \label{fig:dataplot}
\end{figure}

Fig. \ref{fig:dataplot} shows that the effective carrier lifetime $\tau_{\rm{eff}}(t)$ decreased over the year. Fitting the surface component of $\tau_{\rm{eff}}$ to the amphoteric model of Olibet \textit{et al.} \cite{Olibet2007} showed that the defect density at the interface $N(t)$ increased over the year, while $Q(t)$ remained constant. Motivated by these experiments, we developed a robust theoretical framework to simulate,  to model, and  to analyze the time evolution of the interface defect density $N(t)$.

 \subsection*{Si HJ Stacks: Structure, Energies and Defects}
As mentioned earlier, in a previous paper we simulated the degradation dynamics in simplified, Si-only c-Si/a-Si stacks. \cite{soldeg} In that paper, we used the machine learning-based GAP for the Si-Si interatomic potential. However, a crucial ingredient in the experimental samples, hydrogen, was not included in that work. Naturally, a quantitative analysis must include hydrogen. In order to perform high precision modeling, we needed a Si-H interatomic potential with a precision comparable to the Si-Si GAP. Since such a potential was not available, we used machine learning to train and construct the Si-H GAP \textit{de novo}. Here we describe our methods only in bullet points, and refer to the Methods section for details.

First, we developed the Si-H GAP using ML, as reported very recently. \cite{SiHGAP} Most importantly, MD simulations performed with the Si-H GAP were able to reproduce DFT energies with a 4 meV per atom precision. The Si-H GAP enabled us to simulate a-Si:H samples with close to 5,000 atoms, or hundreds of average-sized samples in parallel. Such scales were necessary for the high quality determination of the distributions of key descriptors, needed to account for the measured experimental data.

Next, we used Si-H GAP-powered MD simulations to create separate c-Si and a-Si:H slabs, by validated quenching protocols. 

Then, we merged the c-Si and a-Si:H slabs into stacks using the same Si-H GAP MD, wherein their densities were matched at the interface. We created 60 interfaces.

\begin{figure}[ht]
    \includegraphics[width=0.95\linewidth]{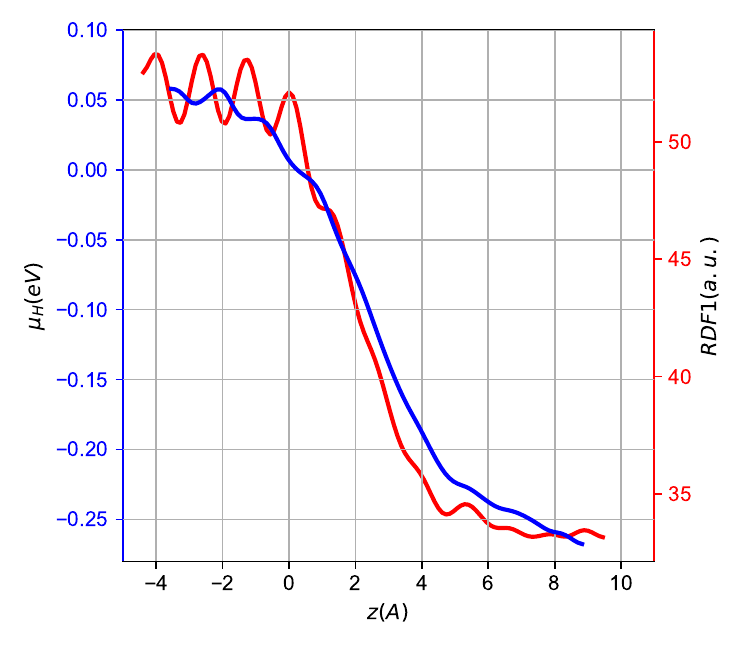}
     \caption{\textbf{Silicon density and hydrogen chemical potential at the c-Si/a-Si:H interface.} The hydrogen chemical potential $\mu_H(z)$ (\textcolor{blue}{blue}), and the Si density as indicated by the first peak height of the radial distribution function $\rm{RDF}1(z)$ (\textcolor{dred}{red}) across the interface of the Si HJ are shown. $\mu_H(z)$ tracks the Si density remarkably closely, making is plausible that the Si density gradient is the cause of the H chemical potential gradient.}
    \label{fig:VH-RDF1}
\end{figure}

Finally, we determined $\mu_H(z)$, the position-dependent energy of adding a hydrogen atom to the stacks at an interstitial location a distance $z$ from the interface. The result of 25,000 such calculations is presented in Fig. \ref{fig:VH-RDF1}, which shows our first central finding: We observed that $\mu_H(z)$ exhibited a gradient across the interface. This energy gradient means that mobile hydrogen atoms at the interface experience a force that causes them to drift away from the interface into the a-Si:H layer. This hydrogen drift leaves behind unpassivated Si DB, and thus degrades the cell performance. In the remainder of this paper, we focus on this discovery as the most promising mechanism of the experimentally measured increase of the defect density in Fig. \ref{fig:dataplot}.

Next, we explored several different possible drivers of this energy gradient. Eventually, we established that the Si density, as characterized via the Radial Distribution Function (RDF), exhibits a very analogous gradient across the interface, see Fig. \ref{fig:VH-RDF1}, and is therefore the most likely driver of the hydrogen chemical potential gradient.

We then investigated the electronic properties of the hydrogen-induced defects and established that in 94\% of the cases, departing hydrogens create localized dangling bonds, and in 6\% of the cases they leave behind delocalized electronic states. Moreover, we showed that the average  Löwdin charge of the generated dangling bonds is -0.1 e, and thus they can be  properly identified as the neutral defects seen in our experiment.

\subsection*{Energy Barrier Distributions Connecting Femtoseconds to Gigaseconds}
We connected the femtosecond time scale of MD simulations to the gigasecond time scale of the experiments by adapting the Nudged Elastic Band Method (NEB) to determine the energy barriers that control the hydrogen dynamics. We found that the key processes were hydrogen breaking free from Si-H bonds, and hydrogen hopping between interstitial sites. We computed the energy barriers for about 650 initial-final state pairs for the breaking of Si-H bonds, and for more than 2000 initial-final state pairs for interstitial hopping, for both the forward and backward directions.

\begin{figure}
    \centering
    \includegraphics[width = 1.0\linewidth]{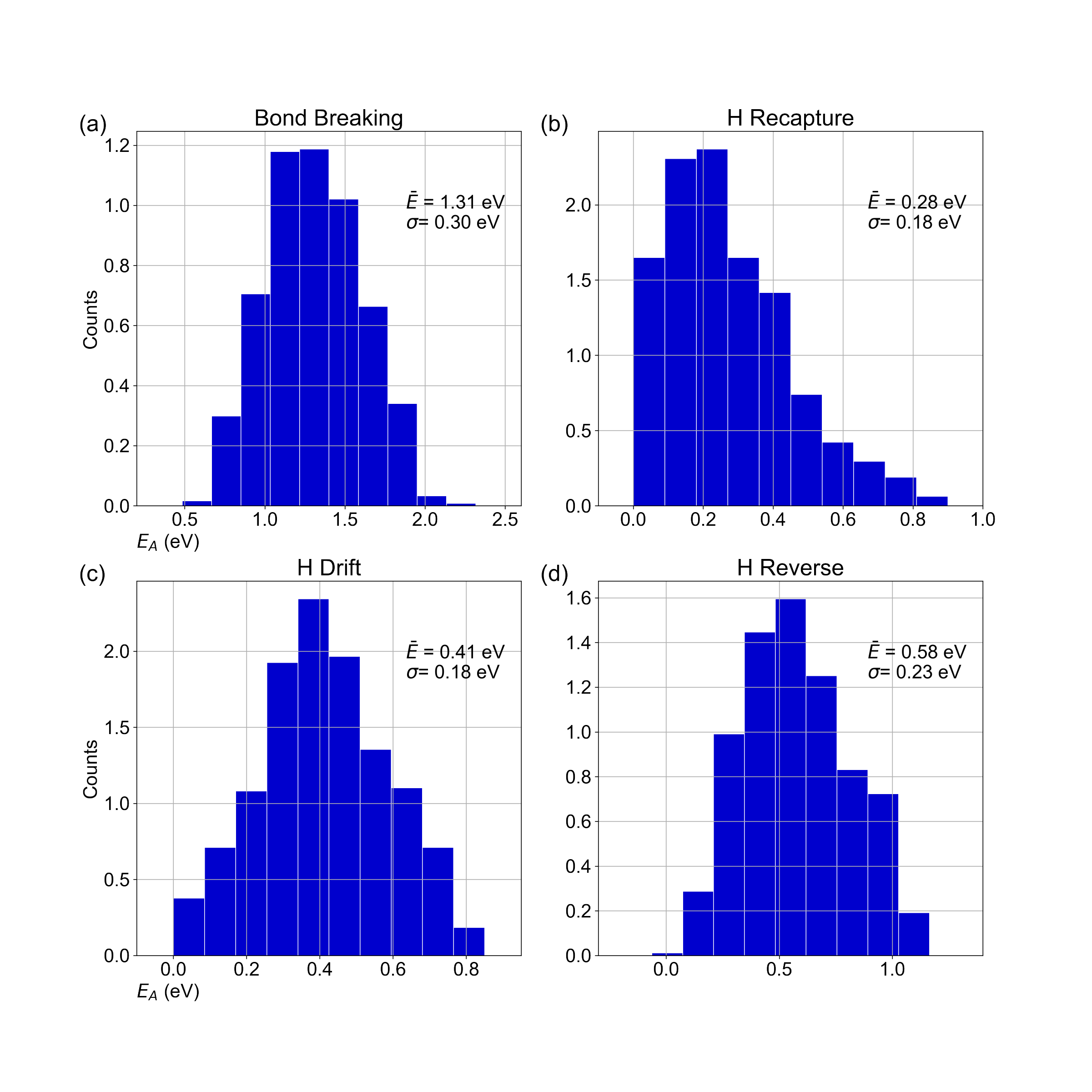}
     \caption{\textbf{Energy barrier distributions that control hydrogen dynamics.} The distribution, mean value and standard deviation of the energy barriers are shown for the four key processes that control the activated thermal motion of hydrogen at the c-Si/a-Si:H interface: \textbf{a}, Freeing a hydrogen atom by breaking a Si-H bond at interface; \textbf{b}, Recapture of the freed H atom back to the Si-H bond; \textbf{c}, Activated drift of H atom away from the interface; \textbf{d}, Reverse drift toward interface.}
    \label{fig:barriers}
\end{figure}

Fig. \ref{fig:barriers} shows our second central finding: The energy barriers for the four processes (bond breaking and interstitial hopping, to and from the interface) all have wide distributions. The computed mean energy barriers of these processes, $\rm{E_{bond}}$=1.31$\pm$0.03 eV, $\rm{E_{inter}}$=0.49$\pm$0.08 eV, were consistent with previous experimental and computational values: $\rm{E_{bond}}$=1.2-1.4 eV, $\rm{E_{inter}}$= 0.48-0.50 eV,\cite{santos_johnson1993,biswas,herring,van_de_walle_street_1995,ww} $\rm{E_{inter}}$ being the average of $\rm{E_{drift}}$ and $\rm{E_{reverse}}$.

In an important distinction, however, most previous studies reported only single values or narrow distributions ($\sigma \approx 0.025 ~\rm{eV}$) for these barriers. In contrast, we found that the width of these distributions were quite large, with $\sigma$ in the range of 0.20 eV-0.30 eV. The substantial width of the energy barrier distributions is important because at room temperatures, activated processes over the mean bond-breaking barrier of 1.31 eV are frozen during the 1 week to 10 years time scale (using attempt frequencies in the $10^{13}$ Hz range from FTIR vibrational data), and thus could not explain the defect density measurements in Fig. 1. Only barriers in the 1.1-1.2 eV range are active in the experimental time range of 1 week-1 year. Notably, such barriers are present in our simulation results.

We also computed the interstitial energy barriers: $\rm{E_{drift}}$ controlling the drift away from the interface into the a-Si:H layer, and $\rm{E_{reverse}}$ controlling the drift back form the a-Si:H layer toward the interface. Most previous papers did not recognize any difference between these processes and reported a single barrier value of $\rm{E_{inter}}=0.50 ~\rm{eV}$. \cite{herring,ww,santos_johnson1993,biswas} However, our first main result, the existence of the  hydrogen chemical potential  gradient, resulted in a bias between these barriers: $\rm{E_{drift}}=0.41  ~\rm{eV}$, and $\rm{E_{reverse}}=0.58 ~\rm{eV}$. While the average of these two barriers, $\rm{E_{inter}}=0.49 ~\rm{eV}$, remained in agreement with the previously published value of 0.50 eV, the bias between them presents a physical explanation for the drift of hydrogen away from the interface. 

Furthermore, we found that $\rm{E_{recap}}$, the barrier that controls the recapture of a hydrogen atom from an interstitial position by a Si dangling bond, was centered at 0.28 eV. Published values identified this barrier also as 0.50 eV. \cite{santos_johnson1993} The difference in $\rm{E_{recap}}$ from the other interstitial barriers is a result of the local low Si density region near a dangling bond. 

\subsection*{Degradation Dynamics} 
Determining the energy barriers and their distributions created the foundation on which we next built the analysis of the dynamics of hydrogen drift. First, we simulated the hydrogen drift from the interface across the first 4-16 local energy minima along the z perpendicular direction, by both  Accelerated Superbasin Kinetic Monte Carlo methods,\cite{soldeg} and as a series of thermally activated jumps, a representation of the energy landscape used in these simulations is shown in Supplementary Figure 2. The results of these simulations illuminated the dramatic effect of the  hydrogen chemical potential  gradient. The bias of the energy barriers relative to $k_{B}T$, $\frac{\rm{E_{drift}} - \rm{E_{reverse}}}{k_{B}T} \approx 6 $ made the probability of a hydrogen returning to the interface less than 0.1\%. This very small return probability validated reducing the dynamics to the following three hydrogen states: H bonded to a Si at the interface, H at an interface-adjacent interstitial site, and H in the bulk of the a-Si layer. The three energy barriers controlling the dynamics between these three states are the Si-H bond-breaking $\rm{E_{bond}}$, the bond-recapture barrier $\rm{E_{recap}}$, and the barrier against forward drift into the a-Si:H layer $\rm{E_{drift}}$, for validation of the Three Barrier Model see Supplementary Figure 3.

\begin{figure}
    \centering
    \includegraphics[width = 1.0\linewidth]{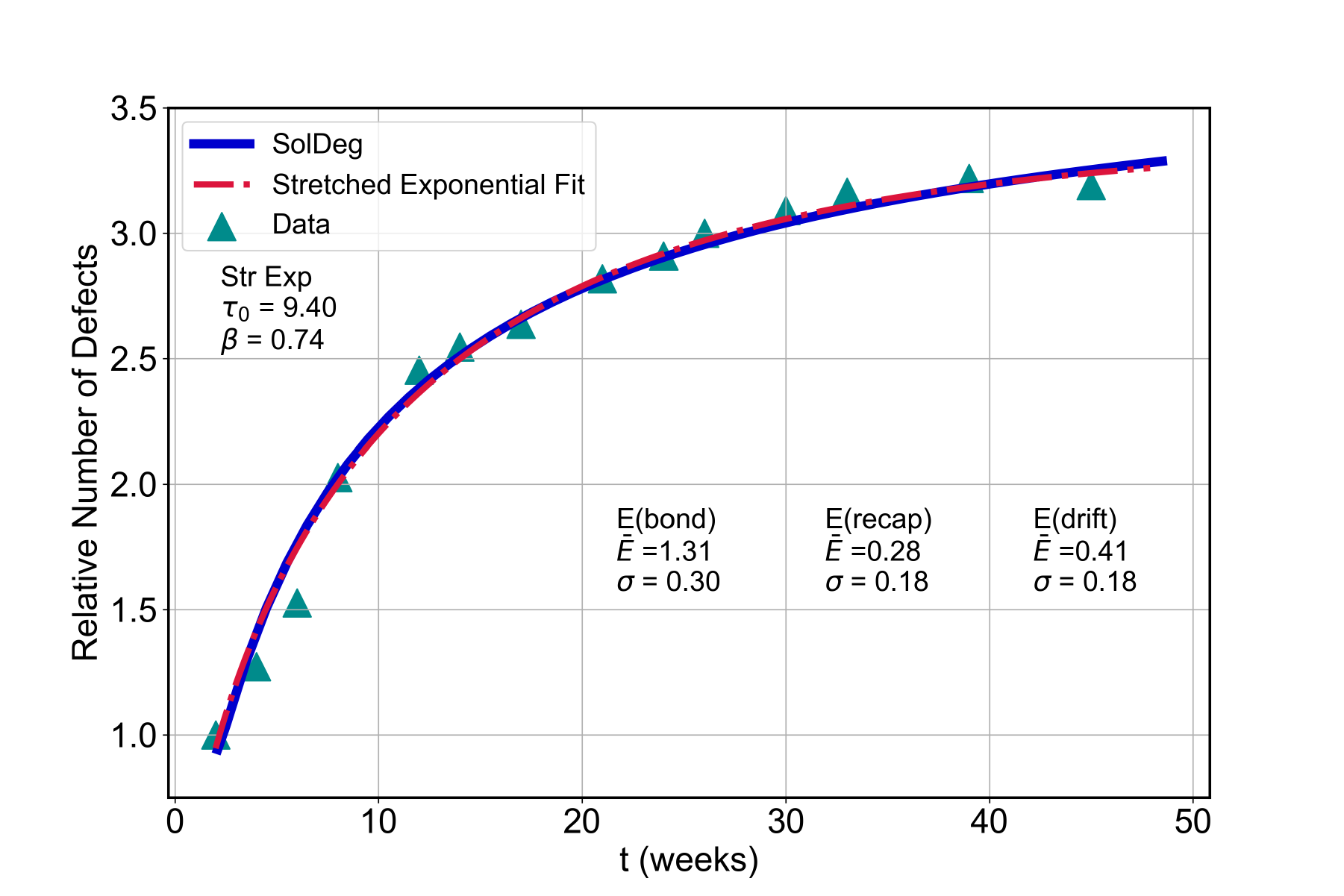}
     \caption{\textbf{Defect density $N(t)$ determined by SolDeg.} The \textcolor{blue}{blue line} shows our SolDeg-simulated defect density as determined by the expectation value of the solution of the three barrier model using the computed barrier distributions in Fig. \ref{fig:barriers}. The experimentally measured $N(t)$ is shown with the \textcolor{trq}{turquoise triangles}. The SolDeg-simulated \textcolor{blue}{blue line} is not a fit: only the overall scale factor was selected to match the measured data. Also shown: stretched exponential fit to the experimental data, with \textcolor{dred}{red dashed line}.}
    \label{fig:da_tbm_strexp}
\end{figure}

We constructed the following set of rate equations for the three barrier model:
\begin{equation}
    \der{\mathbf{H}}{t} = \mathbf{M} \mathbf{H}
\end{equation}
where the $\mathbf{H}$ vector denotes the hydrogen densities in the three model states, and
\begin{equation}
    \mathbf{M}=
\begin{pmatrix}
-k_1 & k_2 & 0\\
k_1 & -(k_2 + k_3) & 0\\
0 & k_3 & 0
\end{pmatrix}
\label{eqn:Mat}
\end{equation}

denotes the matrix of thermally activated rate constants across the above determined energy barriers. We solve this Eq. \ref{eqn:Mat} for the hydrogen density at the interface $H_i(t)$ as:
\begingroup
\small
\begin{equation}
    H_i(t) = H_i(0) e^{-\frac{\alpha t}{2}} \bigg( \rm{cosh} \big( \frac{\beta t}{2} \big) + \frac{\alpha-2k_1}{\beta}
    \rm{sinh} \big( \frac{\beta t}{2} \big) \bigg)
    \label{eq:HI}
\end{equation}
\endgroup

where where $\alpha \equiv k_1 + k_2 + k_3$, and $\beta \equiv \sqrt{\alpha^2 - 4k_1k_3}$. As we established before, when hydrogens leave the interface, they leave behind dangling bonds and thus increase the defect density as $\Delta N_i(t) = - \Delta H_i(t)$. We determined the total defect density $N(t)$ by averaging $N_i(t)$ over the distributions of the three rate constants, which we determined from the barrier distributions computed using the NEB method with our Si-H GAP. From here on, we extend the scope of the SolDeg simulator of Ref. \cite{soldeg} to refer to the here-described multiscale, hierarchical combination of simulations and analytical methods. Fig. \ref{fig:da_tbm_strexp} shows our SolDeg-simulated $N(t)$ alongside the experimentally determined $N(t)$ from Fig. \ref{fig:dataplot}.

The central result of our paper is the remarkable correspondence of the measured and the SolDeg-simulated defect density $N(t)$, see Fig. \ref{fig:da_tbm_strexp}. This compelling correspondence is validation that our SolDeg simulator provides a quantitatively reliable description of the hydrogen-induced degradation dynamics of heterojunction systems. 

Fig. \ref{fig:da_tbm_strexp} further shows that the measured and simulated defect densities can also be described very well with a stretched exponential form, which is a hallmark of relaxation in disordered systems with broad barrier distributions.

The time-dependence of the defect density $N(t)$ has direct experimentally measurable consequences. The increasing defect density makes an additive contribution to the inverse career lifetime $1/\tau$, causing it to decrease over time. Further, following the models of Green \cite{green-ThirdGen} and Olibet \textit{et al.} \cite{Olibet2007}, or Blank \textit{et al.} \cite{Kirchartz2017}, $\voc$ is another quantity directly effected by the defect density $N(t)$ via
\begin{equation}
    \voc(t)=C- \frac{k_{\rm{B}} T}{q} \rm{ln} (N(t)),
    \label{eqn:voct}
\end{equation}

where $C$ is a constant and $q$ is the elementary charge, for a full derivation see Supplementary Equations [4-13] . This relation connects our SolDeg simulation results to the commercially relevant degradation of $V_{oc}(t)$ in Si HJ modules. Additional utility and relevance of our SolDeg platform can be appreciated, e.g., from the fact that the expression for $V_{oc}(t)$, constructed by inserting a stretched exponential $N(t)$ into Eq. (\ref{eqn:voct}), fits the experimentally measured \cite{NRELdegradation} degradation of $V_{oc}(t)$ in fielded modules over the entire measurement range of 8 years, see Supplementary Figure 1. 

\subsection*{Reversed Gradient NoDeg Stacks}  
Having gathered the above insights into the physics of the degradation of existing Si HJ cells, we developed a proposition on how to slow down this degradation. Since the Si density gradient caused the hydrogen drift from the interface, we suspected that reversing this density gradient would do the opposite; it would pin the hydrogen atoms to the interface. To test this idea, we changed the c-Si to a-Si:H slab-merging protocol to create a Si density that was lower at the interface and increased into the a-Si:H layer.

\begin{figure}[ht]
    \includegraphics[width=0.95\linewidth]{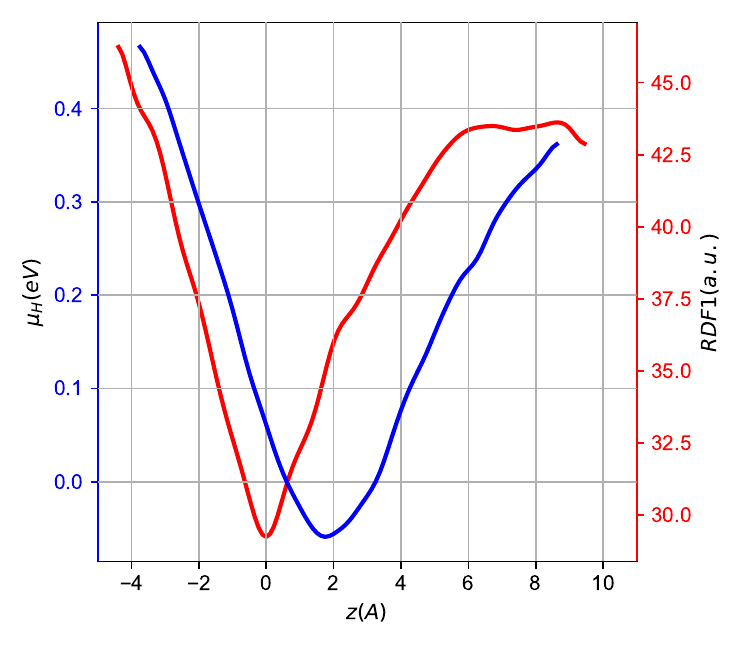}
     \caption{\textbf{Hydrogen chemical potential for a reverse gradient Si HJ interface.} The reversed-gradient stacks were constructed so that the Si density (\textcolor{dred}{red}), as described by the RDF1(z), exhibited a minimum at the interface. The Si density minimum induced a corresponding minimum in $\mu_H(z)$ (\textcolor{blue}{blue}). The reversal of the $\mu_H(z)$ gradient induced a force on the mobile H atoms toward the interface.}
    \label{fig:VH-reverse}
\end{figure}


Fig. \ref{fig:VH-reverse} shows that in these reversed-gradient stacks, the hydrogen atoms feel a force that pins and localizes them to the interface, thereby ensuring that the defect generation, and thus cell degradation by hydrogen drift is suppressed. From here on we refer to heterojunction cells which are fabricated with a Si density minimum at the interface as NoDeg cells.

\begin{figure}
    \centering
    \includegraphics[width = 1.0\linewidth]{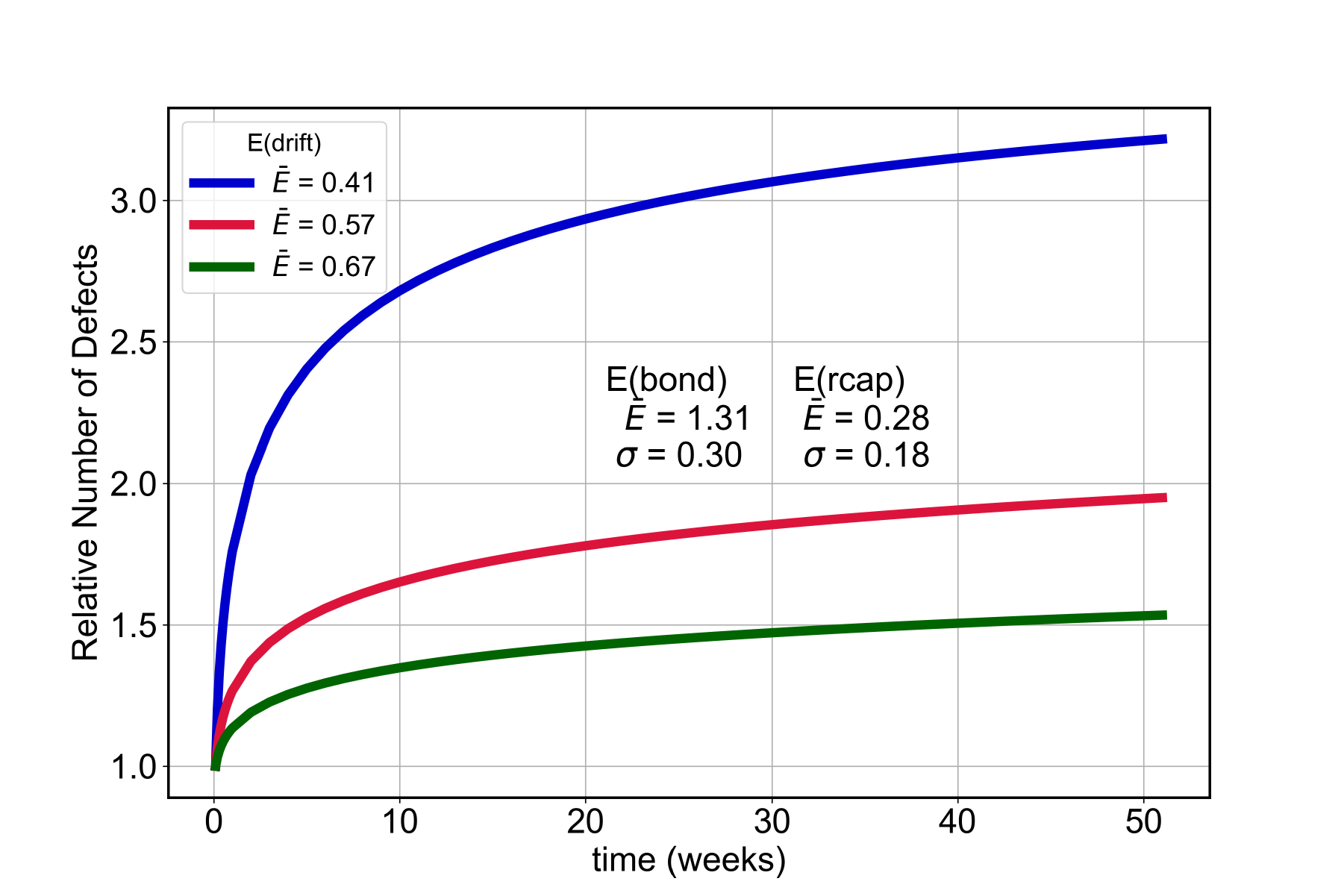}
     \caption{\textbf{Defect growth over time with different $\mu_H(z)$ gradients.}
     SolDeg-simulated results of the interface defect density $N(t)$ with mean values of $\rm{E_{Drift}}$ representing a $\mu_H(z)$ gradient pointing away from the interface - \textcolor{blue}{blue}; or a reversed gradient pointing towards the interface: moderate gradient - \textcolor{dred}{red}, stronger gradient - \textcolor{dgreen}{green}, comparable to the magnitude of the original gradient (\textcolor{blue}{blue}). The defect density $N(t)$ at 1 year was suppressed by 80\% in reversed gradient NoDeg cells. }
    \label{fig:NoDeg}
\end{figure}

We repeated our SolDeg simulation of the hydrogen-induced degradation in these NoDeg stacks. The key difference was that the $\rm{E_{Drift}}$ barrier controlling the forward drift increased as the reversed Si density gradient increased. Fig. \ref{fig:NoDeg} shows the remarkable result of our SolDeg simulations of the NoDeg cells. 

Most remarkably, in these NoDeg cells, after one year the defect density $N(t)$ was reduced relative to the original stacks by about 80\%. Here we recall that the original experiment that motivated our work reported that the open circuit voltage of Si HJ cells, $\voc$ decreased at a rate of 0.5\% per yr. The above simulation of the NoDeg cells suggests that by reversing the Si density gradient at the interface, the $\voc$ degradation rate could be suppressed to a very encouraging 0.1\% per yr. As a practical matter, the Si density and its gradient can be tuned in a wide range by suitably modifying deposition temperatures, partial pressures of the SiH4, H2, and possibly Si2H6 gases, and other deposition parameters. 

After the completion of our work, we learned that some experimental groups fabricated and characterized an analog of the proposed reverse gradient heterojunction cells. These groups created a bilayer architecture of the a-Si:H layer: They deposited a thin underdense a-Si:H on the c-Si, followed by the deposition of a regular a-Si:H layer. All groups measuring heterojunction cells with such bilayer architecture demonstrated the improvement of surface passivation. Lee \textit{et al.}. reported $\voc$ improvements of 10-20 mV when comparing bilayer cells with 5 nm of low density a-Si:H, plus 5 nm regular density a-Si:H layers with cells with a 10 nm a-Si:H layer. \cite{Lee2014} Smets \textit{et al.}. reported a 10 mV $\voc$ increase for a 1nm+8nm bilayer design relative to a 9nm single layer. \cite{ZhaoSmets} Sai \textit{et al.}. reported a 30 mV increase for a 2nm+8nm bilayer relative to a 10 nm single layer. \cite{Sai-2018} Other groups also reported improvements in passivation, fill factor, or $\voc$. \cite{Duan2022,RU-2020,Zhang-2017,Liu-2016}

The above groups did not perform systematic degradation studies. However, we speculate that the reported $\voc$ improvements could have been the consequence of suppressed degradation that took place from the moment of fabrication to the time of the $\voc$ measurements, days or weeks later.

The cited bilayer papers already established that fabricating HJ cells with underdense-dense a-Si:H bilayers improved the passivation. Beyond these, our NoDeg simulations propose that even greater passivation improvements can be achieved by depositing the a-Si:H layer with a continuous reversed Si density gradient. Of course, other physical processes, like the formation of higher hydrides and morphological variations could have played a role in the measured changes of the passivation, and their roles should be explored.

\subsection*{Conclusion}

This paper reported a comprehensive study of the performance degradation of c-Si/a-Si:H heterojunction stacks. We carried out a yearlong experimental study of the performance degradation of these stacks and determined the time dependence of their interface defect density. Then we developed SolDeg, a multiscale, hierarchical simulator, to analyze this degradation. First, we used Machine Learning to construct the most accurate Si-H interatomic Gaussian Approximation Potential GAP that reproduced DFT energies with a 4 meV per atom precision. Then we used this Si-H GAP to carry out LAMMPS Molecular Dynamics simulations to construct c-Si/a-Si:H heterojunction stack interfaces. We discovered that the hydrogen chemical potential developed a gradient across the interface, which forced the hydrogen atoms to drift away from the interface. The departing hydrogens left behind dangling bond defects that acted as recombination centers for the charge carriers of the c-Si wafer. We implemented the Nudged Elastic Band method to determine the barriers that control the hydrogen dynamics, and their distributions. We identified the crucial processes as the breaking of the Si-H bonds and the drift between interstitial states. We determined that the barrier distributions were characterized with remarkably large widths of about 0.30 eV. We constructed and validated a simplified model of the dynamics of the hydrogen-induced defect generation across these barriers. We found a compelling, quantitative correspondence between the calculated and the experimentally measured defect generation dynamics,  as well as a promising connection to experimentally measured $V_{oc}(t)$ degradation in similar fielded modules out to 8 years. Finally, we built on this analysis to propose a reversed Si-density gradient architecture for the a-Si:H layer of heterojunction cells to suppress the interface defect generation. In such NoDeg cells the initial $\voc$ degradation rate may be reduced from 0.5\% per yr to 0.1\% per yr.

\section*{Methods}
 
\subsection*{Experimental Degradation Study}  
For the lifetime measurements, c-Si wafers were bifacially coated with a-Si:H(i). The wafers were double-side polished float zone (FZ) quality n-type c-Si wafers with (100) crystal orientation, $2.5 \Omega cm$ resistivity, and thickness of $\approx 275 \mu m$. The wafers were rigorously cleaned in Piranha $ ( \rm{H_2 SO_4} \colon \rm{H_2 O_2} \vert 4 \colon 1 ) $ and RCA-b $ ( \rm{H_2O} \colon \rm{HCl} \colon \rm{H_2O_2} \vert 6 \colon 1 \colon 1 ) $ solutions.  An important note on crystal orientation: deWolf \textit{et al.}. showed that on long time scales the time dependence of recombination lifetime is independent of the orientation of the crystal axes. Therefore, we believe our results are relevant for HJ cells which use c-Si wafers with (111) orientation at the interface.

Wafers were formed with four different thicknesses W in the range of $160 - 260 \mu$ by etching them over different times in an HNA mixture $(\rm{HF} \colon \rm{HNO_3} \colon \rm{CH_3COOH} \vert 10\colon 73 \colon 17)$, ending with a dip in a buffered oxide $(\rm{HF} \colon \rm{H_2O} \vert 10 \colon 1)$ etch solution, for specific wafer cleaning and etching protocols see Supplementary Methods 1. Subsequently, 50 nm of a-Si:(i) was deposited on both sides of the wafers using plasma-enhanced chemical vapor deposition (PECVD) technique in Octopus I tool from INDEOtec S.A. The effective minority carrier lifetime was measured with the WCT-120TS tool from Sinton Instruments, lowering the temperature from $230^o$ C to $30^o$ C.

The deposited a-Si:H(i) films were characterized with Fourier transform infrared spectroscopy (FTIR) using Nicolet 6700 spectrometer from Thermo Electron to determine the microstructure and hydrogen content of the films. A M2000 ellipsometer from JA Woollam was employed to characterize the thickness, bandgap (Eg), refractive index, and degree of crystallinity in the a-Si:H(i) layer. Ellipsometric spectra were collected at multiple incident angles of 65, 70 and 75 degrees in reflection mode. The resulting spectra were fitted with a single Tauc-Lorentz oscillator and yielded a gap of 1.68 eV.

\subsection*{Development of the Si-H GAP Using ML}  
\begin{figure}[ht]
    \includegraphics[width=\linewidth]{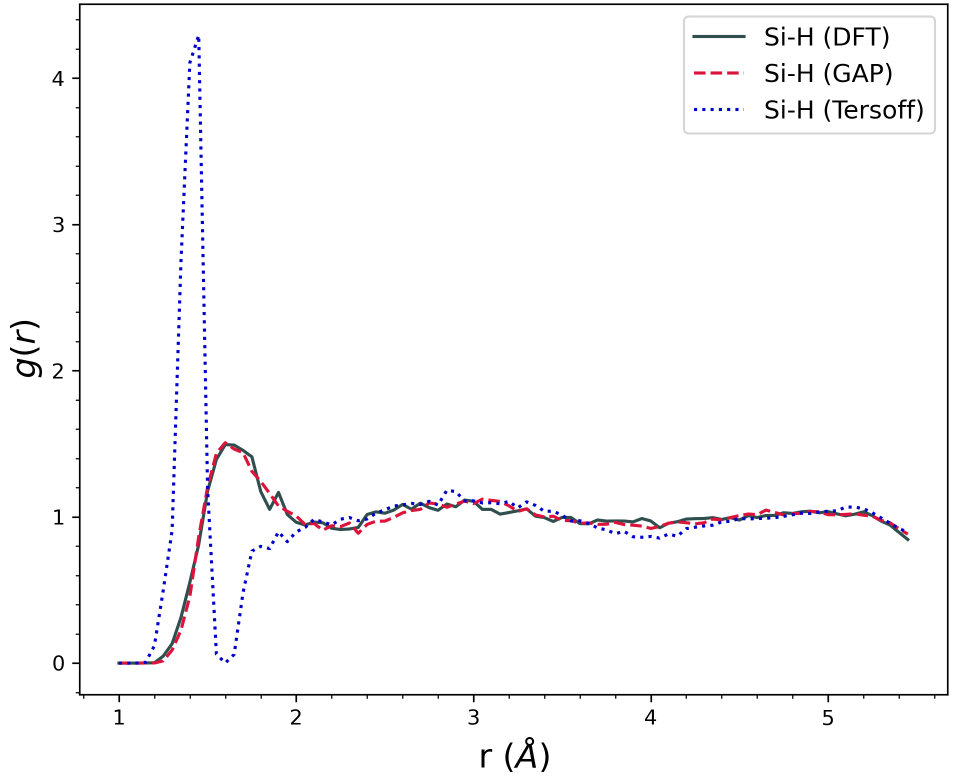}
     \caption{\textbf{Si-H pair correlation function in liquid Si:H.} The liquid Si-H RDF computed with the Si-H GAP and with DFT are barely distinguishable, while both differ from the Tersoff-computed RDF markedly. These results further demonstrate the quality of ML based potentials such as the Si-H GAP.}
    \label{fig:RDF}
\end{figure}

The Si-H GAP was developed by simulating Si samples with 12\% and 15\% hydrogen concentrations. We performed 27 rounds of training of the Si-H GAP on a wide variety of relevant Si:H structures, including amorphous and liquid Si, and several defect and interface structures. The Si-H GAP was validated via several key descriptors. Specifically, our training successfully reduced the difference between the Si-H GAP energy-per-atom and the DFT energy-per-atom to below 4 meV per atom. Given that the differences of the energy-per-atom calculated by different DFT methods, as well as the differences between DFT vs. experiments repeatedly exceed this value, the precision of Si-H GAP we achieved is quite remarkable. Further, the differences of the force differentials between the Si-H GAP and DFT were reduced by about 40\% in the course of the training. For further context, the energy-per-atom, force fields, and stress fields were also substantially reduced compared to the results of non-GAP interatomic potentials, such as the Tersoff potential. \cite{SiHGAP} In addition, the Radial Distribution Function (RDF), shown in Fig. \ref{fig:RDF}, and the Bond Angle Distribution functions were also calculated by our Si-H GAP.

\begin{figure}[ht]
    \includegraphics[width=\linewidth]{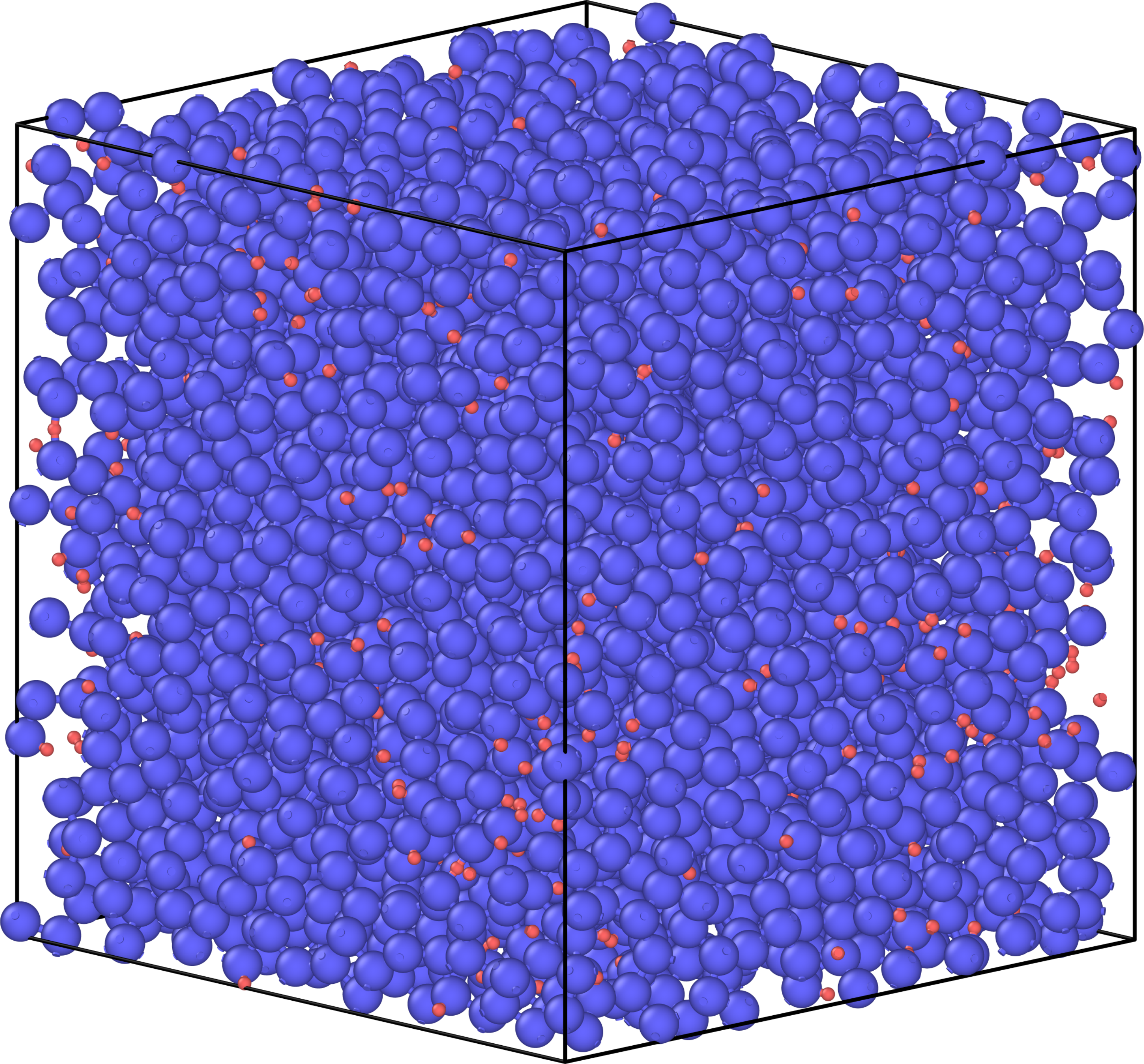}
     \caption{\textbf{a-Si:H supercell, with 4096 Si atoms and 558 H atoms.} Simulations of this size are utterly inaccessible by DFT calculations. The Si-H GAP makes it possible to create exceptionally large samples, and thus to determine quantities that would be missed by smaller DFT simulations, an example of which is the statistical distribution of void sizes.}
    \label{fig:supercell}
\end{figure}

Finally, we demonstrated the remarkable benefit of having constructed the Si-H GAP by creating a hydrogenated amorphous slab of 4,096 Si atoms and 558 H atoms with our Si-H GAP, see Fig. \ref{fig:supercell}. Such simulations are utterly inaccessible by DFT calculations. The Si-H GAP makes it possible to create exceptionally large samples, and thus to determine quantities inaccessible for smaller DFT simulations, an example of which is the statistical distribution of void sizes.  

\subsection*{Creation of c-Si/a-Si:H Stacks}
 Armed with the freshly-developed Si-H GAP potential, we proceeded to create hydrogenated c-Si/a-Si:H stacks. For our work, we used Molecular Dynamics simulations, in particular the LAMMPS package. The c-Si slabs contained 162 Si atoms. The a-Si slabs were created by starting with a 216-atom c-Si slab, and inserting into it 28 hydrogen atoms at random locations, creating an a-Si:H layer with 13 at.\% H, a typical value for the passivating layer of Si HJs. Next, these hydrogen-doped Si slabs were heated to 1,800K by using the Si-H GAP to form liquid Si:H. The liquid Si:H was subsequently re-solidified by quenching to 1,500K at a rate of $10^{13} Ks^{-1}$, and then equilibrated at 1500K for 100 ps. This solid Si:H was then further quenched down to 500K at a rate of $10^{12} Ks^{-1}$, following protocols optimized in previous studies. \cite{PRX,Ishimaru1997,Stich1991,Jarolimek2009} The first quench was performed in the constant-volume and variable-pressure (NVT) ensemble, while the second quench was performed in the variable-volume and constant-pressure (NPT) ensemble with fixed x and y cell-dimensions to match the dimensions of the c-Si slab in the later steps. Both quenches used a Nose-Hoover thermostat and barostat. We minimized the structural energy using a Hessian-free truncated Newton (HFTN) algorithm to relax all atomic positions into their local minima. These relaxed a-Si:H structures were further optimized with DFT, for DFT methods and parameters see Supplementary Methods 2. 
 
\begin{figure}[h]
    \includegraphics[width=0.95\linewidth]{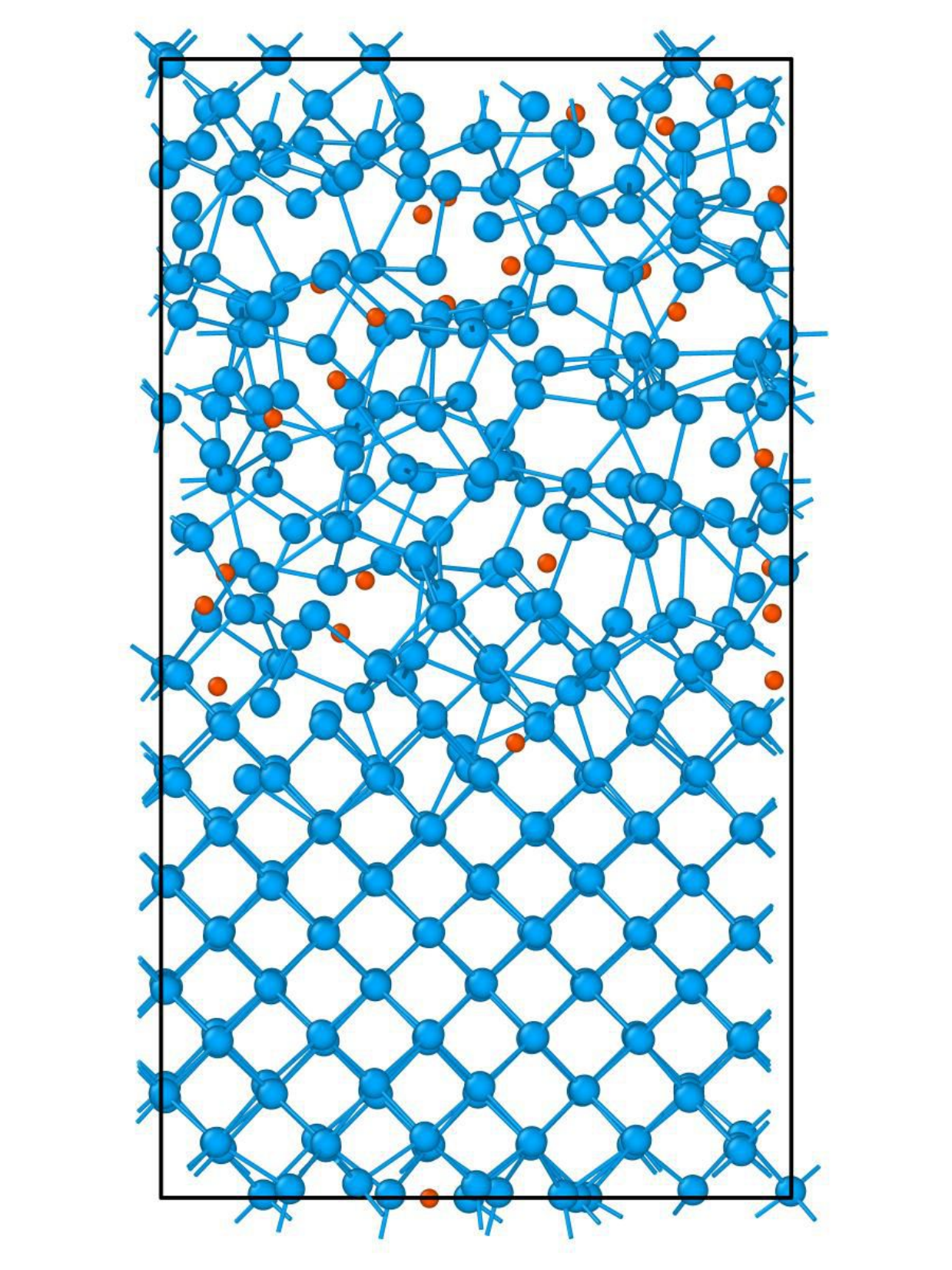}
     \caption{\textbf{c-Si/a-Si:H interface structure.} Shown is a stick and ball model of one of the final interface structures used in our study. The \textcolor{lblue}{blue} atoms represent silicon and the \textcolor{dred}{red} atoms represent hydrogen. }
    \label{fig:merged}
\end{figure}

\begin{figure}[htp]
    \includegraphics[width=0.95\linewidth]{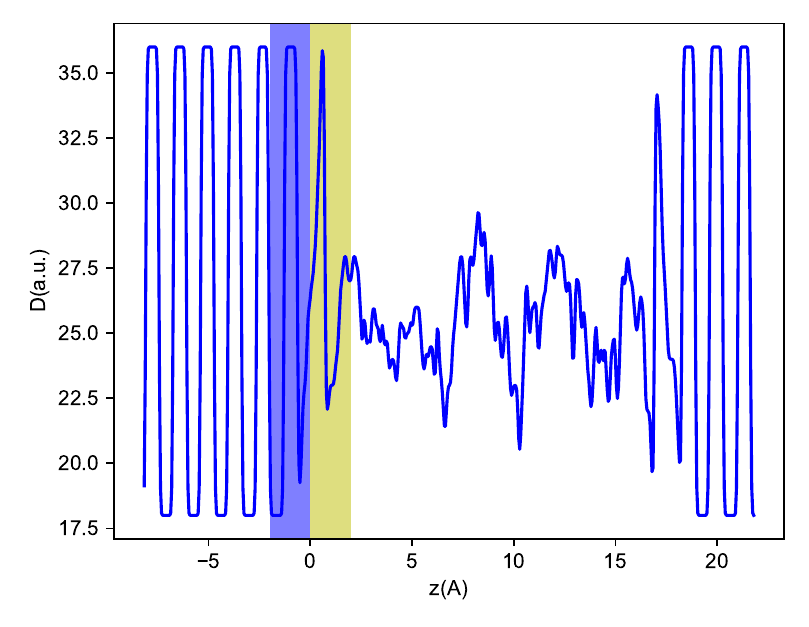}
     \caption{\textbf{Density matching at the left interface.} The z-dependent density of a stack averaged within a 2.5 Å thick sliding cuboid, centered at z and integrated over the lateral $r=(x,y)$ coordinate across the entire lateral extent of the stack. The shaded regions show the 2.5 Å cuboid representing (\textcolor{blue}{blue}) the top of c-Si, and (\textcolor{dyellow}{yellow}) the bottom of the a-Si:H. The peak height in both regions shows the smooth transition of the Si density across the interface, similar to the desired structure of the transition zone passivating layer.}
    \label{fig:DensityMatched}
\end{figure}

In the second stage, the a-Si:H slabs were fused with the c-Si slabs, thus creating an interface and forming the c-Si/a-Si:H heterojunction, see Fig. \ref{fig:merged}. The c-Si/a-Si:H structures will be also referred to as stacks. First, we created a set of stacks by positioning the slabs at a sequence of separations. Next, we determined the z-dependent density D(z) of each stack as the density averaged within a 2.5 Å thick sliding cuboid, centered at z and integrated over the lateral $r=(x,y)$ coordinate across the entire lateral extent of the stack. This density  D(z) is shown for one of the stacks in Fig. \ref{fig:DensityMatched}. Then we selected the stack with the a-Si:H -- c-Si slab separation that caused the density  D(z) to have the smoothest transition from the c-Si slab into the a-Si:H slab. In effect, we matched the density of the top 2.5 Å of the c-Si with that of the bottom 2.5 Å of the a-Si:H, indicated by the blue and yellow rectangles in Fig. \ref{fig:DensityMatched}.

The zero of the z coordinate indicates the interface, i.e. location where the slabs were fused. This method was designed to simulate the commonly used transition zone deposition procedure of a-Si:H(i) layers. The fusing was completed by a final annealing step, where we annealed the fused stacks by heating them to 1500 K, and then quenched them at a rate of $10^{13} Ks^{-1}$. This allowed the distortions and defects generated by the fusing at the interface to diffuse or relax into the bulk of the a-Si:H slab.

\subsection*{Hydrogen Chemical Potential Gradient}
Next, we computed several characteristics of the stacks. We began by calculating the energy $\mu_H(z)$ experienced by a probe hydrogen atom inserted in a stack at a depth $z$. We inserted these probe hydrogen atoms into each stack at 300-500 randomly selected locations $(x,y,z)$ and determined the total energy $\mu_H(x,y,z)$ of the resulting structure. All of the probe hydrogen atoms relaxed into interstitial sites at the center of Si-Si bonds  as almost all dangling bonds in the stacks were passivated by existing hydrogens. The $\mu_H(x,y,z)$ energies were smoothed by averaging over all insertion locations within a 2.5 $\rm{\AA}$ thick sliding cuboid, just like we did with the density  D(z) earlier. Finally, we slid the center of this averaging cubiod in small increments across the entire stack to construct {$\mu_H(z)$}. Carrying out this procedure for the 60 interfaces (2 interfaces per stack), our comprehensive $\mu_H(z)$ is the result of about 25,000 calculations. Fig. \ref{fig:VH-RDF1} shows the so-calculated $\mu_H(z)$  hydrogen chemical potential.

\subsection*{Origin of the Hydrogen Chemical Potential Gradient}
Next, we explored several different possible drivers of this energy gradient. Eventually, we suspected that the Si density and microstructure played a crucial role. To test this hypothesis, we analyzed the radial distribution function  $\rm{RDF}(r,z)$, which was again calculated using a 2.5 Å thick cuboid centered on a $z$ coordinate, as a function of the lateral radius $r = (x,y)$. Specifically, we characterized the local microstructures with the height of the first peak of  $\rm{RDF}(r,z)$ as a function of the lateral radius r, which we denoted by  $\rm{RDF}1(z)$. A high  $\rm{RDF}1(z)$ indicated a high degree of crystallinity and therefore high density, which translated to a low porosity, whereas a low  $\rm{RDF}1(z)$ indicated the opposite. The red line in Fig. \ref{fig:VH-RDF1} shows  $\rm{RDF}1(z)$ as a function of z.  $\rm{RDF}1(z)$ visibly also exhibits a marked gradient across the interface at $z=0$. The correlation between the gradient of the  hydrogen chemical potential  $\mu_H(z)$ and the gradient of  $\rm{RDF}1(z)$ is conspicuous. This makes us confident to conclude that the gradient of the  hydrogen chemical potential  was created by a corresponding gradient of the microstructural crystallinity, and thus density across the interface of the heterjunction, as the stack transitioned from the crystalline Si to the less crystalline, and thus more porous, amorphous Si.

\subsection*{The Nudged Elastic Band Method}  
The experimental study determined the degradation dynamics up to a year. Solar cell manufacturers make guarantees or warranties out to 20-30 years (the latter being equal to about 1 gigasecond). In contrast, the time step of our LAMMPS MD calculations is less than 1 femtosecond. As such, MD calculations have no chance of reaching anywhere near the experimentally relevant time scales. In order to bridge the 24 orders of magnitude chasm between these time scales, we recognize that thermally activated stochastic processes across energy barriers in the $\approx 1 eV$ range are the ones that control the hydrogen dynamics on the slow 1-50 weeks experimental time scale. In order to describe the degradation dynamics on such a slow scale, we must determine the energy barriers the hydrogen atoms face up to the eV scale. (Here we assumed reasonable attempt frequencies of about $10^{13}$ Hz, corresponding to local vibration modes, as determined by Fourier Transform Infrared spectroscopy, or FTIR.)

To compute the energy barriers, we turned to the Nudged Elastic Band (NEB) method. The NEB method is based on constructing multiple replicas of a given system, corresponding to a path from an initial to a final state over a high energy region. In an informative scenario, the initial state is a hydrogen in a local minimum (bound to a Si, or at an interstitial position), the final state is the same hydrogen at another local minimum, and the replicas are intermediate configurations where the hydrogen crosses a high energy region between these minima, tracking both the displacements of the hydrogen and surrounding Si atoms. The NEB method connects these replica configurations by imaginary springs which provide inter-replica forces, $\vb{F_{sp}}$. The total force on the atoms in each replica is the sum of the interatomic forces and inter-replica forces. The NEB method minimizes the total energy of all replicas, and thereby determines the Minimum Energy Path (MEP) between the initial and final configurations. The activation energy of such a process is then determined as the energy difference between the saddle point/maximum energy configuration and the initial configuration. Specifically, we used the climbing  image NEB method,\cite{nebA,cneb} as implemented by LAMMPS, with the Si-H GAP\cite{SiHGAP} interatomic potential and the FIRE\cite{FIRE} damped minimizer, for a complete set of NEB parameters see Supplementary Table 3.

\subsection*{NEB Initial and Final States}

Previous studies have presented a description of hydrogen in a-Si:H, where hydrogen is located in either of two classes: a deep trap, a hydrogen atom bound to an silicon atom; or a shallow trap, a hydrogen at an interstitial Si-Si bond center.\cite{Herring2001,DeWolfSTREXP,santos_johnson1993,KakaliosStreet1987,randwalk} 
Following this model, we organized our NEB calculations into two classes: the energy barriers for the breaking of a Si-H bond and the energy barriers for hopping between interstitial sites. The interstitial hopping barriers were calculated with initial and final states taken from the $\mu_H(z)$ calculations, such that the H atom moves in the z direction by approximately one Si-Si bond length, away from the interface. The bond-breaking initial and final states were determined by an algorithm that scanned the heterojunction stacks to find all mono-hydride Si-H bonds. We identified the mono-hydride Si-H bonds by the coupled criteria that the hydrogen was separated from a single nearest Si by a bond length of about $1.50\rm{\AA}$, in agreement with established bond length values \cite{herring,TBMD,Tersoff1989,SiHGAP}, while the Si was not bonded to a second hydrogen. It was also confirmed that the hydrogen was not at a bond center by checking for multiple Si neighbors within the cutoff distance of $1.7 \rm{\AA}$. The mono-hydride Si-H were chosen as initial configurations for the NEB analysis. For the final configurations, the hydrogen was moved from this mono-hydride bond to a next nearest neighbor bond-centered position. To remain consistent with established Si-H-Si configurations, the two Si atoms were simultaneously displaced by $0.45 \rm{\AA}$ along the direction of the bond. The energies of the initial and final states were then minimized using the FIRE minimizer to an energy tolerance of $10^{-3} \rm{eV}$ before commencing the NEB analysis. After developing and establishing the validity of this method we proceeded to compute the energy barriers for 657 initial-final state pairs for the breaking of Si-H bonds, and for more than 2000 initial-final state pairs for interstitial hopping. We determined the energy barriers for both the forward and backward direction of these processes.

\subsection*{Electronic Properties of Defects}
We completed the characterization of these stacks by determining some of their electronic properties. We decided to measure the inverse participation ratio (IPR) and the partial charge density in each stack. In short, the IPR characterizes how localized an electron wavefunction is \cite{soldeg}, for a details on the IPR see Supplementary Equations [16-18]. The ideal case of an IPR=1 indicates complete localization of the electron wavefunction on a single atom. Complete delocalization is indicated by the IPR value of the order of ($\rm{\frac{1}{N}}$), in our case, 1/406. To adopt a pragmatic convention, we took IPR values exceeding 0.1 as an indicator that the electronic wavefunction was substantially localized on one or just a few atoms. We have described our IPR method in great detail in a recent paper. \cite{soldeg} Using this IPR method, we determined the change in the number of localized defects when a hydrogen atom broke away from a Si-H bond and moved to an interstitial Si-Si bond center. We computed 100 such processes and found that when a hydrogen broke away from a Si-H bond at the interface, on average it induced 0.94 localized dangling bonds. This justifies us identifying a hydrogen breaking away from a Si-H bond at the interface as a defect-generating event. 

\begin{figure}[h]
    \includegraphics[width=0.95\linewidth]{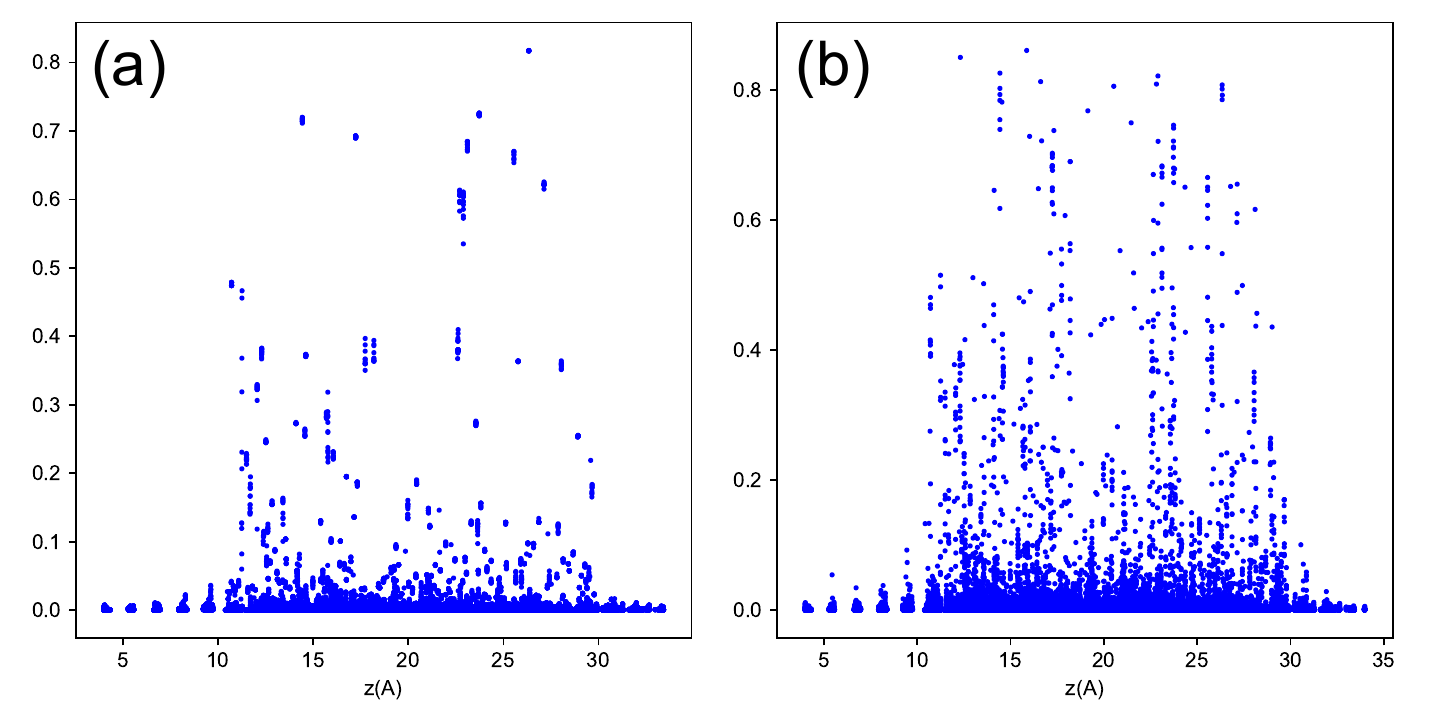}
   \caption{\textbf{Defect characterization at the c-Si/a-Si:H interface.} IPR of atoms at the c-Si/a-Si:H interface and in aSi:H regions \textbf{a}, before and \textbf{b}, after a hydrogen atom is taken out of a dangling bond and placed at a Si-Si bond center. }
   \label{IPR}
\end{figure}

The statistics of these defect-generating events were found to be surprisingly broad numerically and spatially. Several bond-breaking events generated 2-4 localized dangling bonds. Also, the IPR value changed not only on the center Si, but also on atoms several interatomic spacings away. \cite{RezaDefects} Both of these facts are consistent with the softness and glassy interconnected dynamics of a-Si. Fig. \ref{IPR}(a) shows the $IPR_{nkj}$s, calculated for all Kohn-Sham orbitals obtained by DFT as a function of their energy for a typical c-Si/a-Si:H stack.
    
For completeness, we also calculated the partial charge of these left-behind dangling bonds. We found that their partial charge was -0.1e $\pm$ 0.1e. As such, it is appropriate to approximately identify these dangling bonds as neutral defects. This brings the theoretical simulation into agreement with the experimental analysis that explained the degradation data in terms of neutral defects. Partial charges were obtained using the Lowdin population analysis, where plane wave wavefunctions are projected on to the atomic orbitals of atoms.\\

\section*{Data availability}
The data that support the findings of this study are available from the authors on reasonable request, see author contributions for specific data sets.

\section*{Code availability}

The Si-H GAP can be found at {\tt https://github.com/dgunruh/Si-H-GAP} and {\tt https://libatoms.github.io/GAP/data.html}.

The GAP suite of programs is freely available for non-commercial use from {\tt http://github.com/libatoms/GAP}.

The Quantum Espresso software package is freely available from {\tt www.quantum-espresso.org}. 

The LAMMPS software package is freely available from {\tt lammps.sandia.gov}. 
\newpage
\bibliography{main}

\section*{Acknowledgements}
We acknowledge useful discussions with Christophe Balliff and Chase Hansen. This material is based upon work supported by the U.S. Department of Energy’s Office of Energy Efficiency and Renewable Energy (EERE) under the Solar Energy Technologies Office Award Numbers DE-EE0008979 and DE-EE0009835. The views expressed herein do not necessarily represent the views of the U.S. Department of Energy or the United States Government. Work performed at the Center for Nanoscale Materials, a U.S. Department of Energy Office of Science User Facility, was supported by the U.S. DOE, Office of Basic Energy Sciences, under Contract No. DE-AC02-06CH11357. This research used resources of the National Energy Research Scientific Computing Center (NERSC), a U.S. Department of Energy Office of Science User Facility located at Lawrence Berkeley National Laboratory, operated under Contract No. DE-AC02-05CH11231 using NERSC award BES-ERCAP m3634.

\section*{Author contributions}
AD developed the three barrier model, analyzed and fitted the experimental data, and played an energetic role in writing the paper. ZZ created the slabs, fused them into stacks, invented the  hydrogen chemical potential analysis, and analyzed the number and charge of the generated defects; his contribution is entirely comparable to AD, and thus both are equally deserving to be first authors. RVM performed the stack-structure optimizations with DFT methods and the IPR analysis. DU developed the sample creation protocols and the Kinetic Monte Carlo simulations. SM performed the degradation experiments. MB guided and oversaw the degradation experiments. SMG contributed to the development of the physical picture and the analysis and provided useful guidance throughout. GTZ conceived the SolDeg method and the physical analysis, directed the entire project, and played a primary role in writing the paper.

\section*{Competing interests}
The authors declare no competing interests.

\section*{Additional information}
The Supplementary Information is not included in the arXiv version of this article.

\end{document}